\documentclass{ab}

\usepackage{graphicx}
\usepackage{natbib}
\usepackage{lscape,rotating}

\def\pak{$PA_{\rm kin}$\,}
\def\degr{\hbox{$^\circ$}}

\def\arcsec{\hbox{$^{\prime\prime}$}}

\begin{document}

\title{Inner Polar Rings and Disks: Observed Properties}

\author{A.V.~Moiseev}
\institute{Special Astrophysical Observatory Russian Academy of Sciences,
Nizhnij Arkhyz, 369167, Russia}

\date{February 13, 2012/Revised: February 20, 2012}
\offprints{A.V.  Moiseev, \email{moisav@gmail.com}}

\titlerunning{Inner Polar Rings and Disks}
\authorrunning{Moiseev}%

\abstract{A list of galaxies with inner regions revealing polar (or
strongly inclined to the main galactic plane) disks and rings is
compiled from the literature data. The list contains 47 galaxies
of all morphological types, from E to Irr. We consider the
statistics of the parameters of polar structures known from
observations. The radii of the majority of them do not exceed
1.5~kpc. The polar structures are equally common in barred and
unbarred galaxies. At the same time, if a galaxy has a bar (or a
triaxial bulge), this leads to the polar disk stabilization --- its axis of rotation usually coincides with the major axis of
the bar. More than two thirds of all considered galaxies reveal
one or another sign of recent interaction or merging. This fact
indicates a direct relation between the external environment and
the presence of an inner polar structure.
}


\maketitle

\section{INTRODUCTION\label{intro}}

When we generally talk about the presence  in a galaxy of
subsystems rotating in mutually orthogonal planes, we mean the
so-called polar ring galaxies (PRG). The central galaxy here
(typically, of an early-type, E/S0) is surrounded by an external
ring or even by a stellar-gaseous disk, sized up to several tens
of kiloparsecs, positioned roughly perpendicular to the galactic
plane. The first studies of the internal kinematics of PRGs date
back to the late 1970s, while their mass study began after the
publication by \citet{Whitmore1990} of a photographic catalog of PRG
candidates. Despite the lack of
detailed data on the dynamics, evolution and history of star
formation in the PRGs, the key issues can now be considered
solved. In the presence of a triaxial  or spheroidal dark halo,
polar orientation is stable with respect to the differential
precession, so that the ring can make a lot of revolutions around
the central galaxy undisturbed. Numerical models demonstrate that
the PRGs are the result of interaction with the surrounding
matter, the moment of rotation of which is perpendicular to the
rotation axis of the galaxy. The basic mechanisms usually
discussed are the capture of matter from the donor galaxy, the
merger of two orthogonally oriented disks, and for the most
massive and extended \mbox{rings---accretion} of gas from the
intergalactic filaments (see   references in 
\citet{Combes06} review).

Despite the relatively rare occurrence among the nearby galaxies,
the phenomenon of external polar rings is widely known. At the
same time, the literature describes the cases of inner polar
rings and disks, usually scaled below one kiloparsec. This
phenomenon is more poorly studied, which may as well be explained
by the lower ``clarity'' of such structures, usually invisible in
the optical images of galaxies in contrast to the ``classical''
PRGs. It requires quite a lot of effort to identify the internal
polar or inclined disk against the bright bulge, and especially to
obtain detailed data on the motions of the gaseous and stellar
components in the central regions of galaxies.

Interestingly, the inner polar  structures (IPS) were known even
before the phenomenon of PRG has been recognized and confirmed.
For instance, during the spectral observations of the Sc galaxy
NGC\,3672 a significant gradient of the line-of-sight velocities
along its minor axis was detected~\citep{Rub1977}.
According to the authors, this indicates that the axis of rotation
of the circumnuclear ($r<350$~pc) gas has a considerable angle
with the axis of rotation of the galaxy. The alternative
explanation proposed---compression of the gaseous disk in the
galactic plane---seems to be less convincing. Later \citet{Bettoni1990}, according to the results of the
long-slit spectroscopy of  NGC\,2217 have shown that within the
central kiloparsec, a disk of ionized gas is warped in such a way
that rotation occurs in the plane, perpendicular to the stellar
disk of the galaxy. Moreover, the axis of rotation of this polar
disk practically coincides with the major axis of the stellar bar
of NGC\,2217. In the following decade, similar kinematically
decoupled structures were found by several authors within the
detailed studies of internal kinematics of other nearby early-type
galaxies. Note, first of all, the researches made by
Olga Sil'chenko and her colleagues at the \mbox{6-m} BTA telescope
of the Special Astrophysical Observatory of Russian Academy of
Sciences (SAO RAS) using the methods of panoramic (3D)
spectroscopy \citep{Silchenko1997,Sil_N7217,Silchenko2002},
as well as the publications related with the group of the University of
Padua \citep{Bertola2000,Pizzella2001}. The
list of galaxies with confirmed IPSs, published in 2003, already
contained 17 objects~\citep{Corsini2003}. In the subsequent
decade, various groups have presented a fairly extensive
observational material, dedicated to the detection and
investigation of such structures. This was much facilitated by the
survey of kinematics and stellar population of nearby early-type
galaxies, performed at the \mbox{4.2-m WHT} telescope, and the
study of chemically decoupled galactic nuclei at the SAO RAS
\mbox{6-m BTA} telescope, performed using the SAURON and MPFS
integral-field spectrographs, respectively. The
study by \citet{SilAfan2004} is also representative. Here, the
authors selected for the MPFS observations eight galaxies, the
optical images of the central regions of which are clearly
revealing the dust lanes, projected onto the nucleus, which argues
in favor of the presence of gas-dust disks, strongly inclined to
the line of sight. The derived velocity fields of ionized gas and
stars have confirmed the existence of internal disks or polar
rings in all the sample objects. \citet{Coccato2004} have demonstrated that in
\mbox{50--60\%} of bright unbarred galaxies the remarkable
gradient of the line-of-sight velocity is observed along the minor
axis, which may partly be explained by the presence of IPSs.
In~\citet{Moiseev2010} we briefly presented a new list of
37~galaxies with IPSs. Despite the fact that the number of such
objects surpasses the number of kinematically confirmed external
polar rings, their nature remains vague. Up to date, there is no
clear self-consistent scenario of their formation, the issues of
stability  of such structures are not resolved either. The views
on the relationship of IPSs with bars of galaxies and their
external environment (the presence of companions, traces of
interaction, etc.) are contradictory \citep{Corsini2003,Moiseev2010}.

This paper presents an updated list of galaxies with inner polar
rings and disks, compiled based on the data published in the
literature, including our own observational data obtained with the
\mbox{6-m} telescope. A large enough number of objects allowed us
to consider some statistical relations in the properties of IPSs.

\section{COMPILATION OF THE LIST}

\subsection{Selection Criteria}

Table~\ref{tab1} presents the main parameters of internal
polar structures, described in the literature. The columns with
respective numbers contain the following data:\\
 (1) the name of the galaxy;\\
(2), (3) its morphological type  according to  the NED/RC3
and its digital code adopted from the HyperLeda
database\footnote{{http://leda.univ-lyon1.fr}}, (\mbox{$T=-2$}
corresponds to S0, \mbox{$T=0$}---to S0a, etc.). For NGC\,7468,
which is clearly  not elliptical (as specified in the LEDA),
\mbox{$T=9$} was adopted according
to~\citet{Shalyapina2004};\\
(4) the distance ($D$) in Mpc in accordance with the HyperLeda;\\
(5), (6) the external  radius of the polar structure in the
angular and linear scales ($r$). In many papers the authors
themselves gave this value. In other cases, we made the estimates
based on the data presented in the original papers: the diagrams
of radial variations of the kinematic axis  position angle (\pak)
or the published velocity fields. Sometimes the problem was
simplified by the fact that all the ionized gas, observed at the
center of the galaxy  belongs to the polar structure. In this
case, instead of looking for a region of the line-of-sight
velocity gradient variation, it was sufficient to estimate the
size of the region, occupied by the gas emission lines (for
example, NGC\,4552, and NGC\,5129 according to the SAURON 
data). For some galaxies, only to the lower limit of this
parameter is  known, limited by the spectrograph field-of-view;\\
 (7), (8) the parameters of orientation of the main galactic disk:
the position angle ($PA_0$) and the inclination to the line of
sight ($i_0$) in degrees  are in most cases taken from the
original papers, and in the remaining cases---according to the
HyperLeda database;\\
 (9) the position angle of the major axis of the bar
($PA_{\rm bar}$), specified in the original papers, in degrees. In
some cases, (NGC\,2768, NGC\,2841,  NGC\,6340, NGC\,7217) we deal
with the ``triaxial bulge'', rather than the contrasting bar. For
NGC\,4548 we list the orientation of the internal triaxial
structure instead of the external bar (in agreement
with~\citet{Silchenko2002}). In some cases, the authors of
the original papers suspected ``hidden triaxiality'' of the inner
regions based on the indirect evidence, but could not specify the
exact position angle (e.g., see
NGC\,3607 in  \citet{AfanaSil2007});\\
(10) the major axis of the circumnuclear structure ($PA_1$)
in degrees. It was given by the authors or estimated by us from
their diagrams of radial variations \pak$(r)$. The value is
missing for some galaxies, the data on the internal kinematics of
which are based only on the long-slit spectroscopy (NGC\,4424,
NGC\,4698, NGC\,4941) or for which the authors have constructed
the spatial model of the internal structure (Arp\,220, NGC\,1068,
NGC\,2855, NGC\,3227, NGC\,7049). For the  NGC\,3368 and NGC\,4111
galaxies, the adopted parameter $PA_1$ significantly differs from
the estimates based on \pak and was determined from the
orientation of the internal dust ring. We believe that here, in
the velocity field along the line of sight, we observe a
superposition of two gaseous subsystems, or the inclined structure
is nonstationary. For NGC\,5014 the parameters of orientation were
determined from the SDSS image, which reveals a narrow blue
ring~\citep{Moiseev_SDSS};\\
(11) the inclination of the internal structure to the
line of sight~($i_1$), in degrees. Evaluated the same way as
$PA_1$, with the same remarks on individual galaxies.
Unfortunately, in many cases it is impossible to estimate this
parameter, we hence  either list the assumed range of values, or
the lower limit when it is clear that the inner disk is
significantly inclined (NGC\,3414, NGC\,7742, etc.). For
NGC\,2787, and NGC\,2911 our own estimate of orientation of the
dust structure is given;\\
(12) the angle of inclination $\Delta i$ of the internal
structure to the plane  of the disk. The sign $^*$ marks the
estimates from the literature, in other cases it was evaluated by
us (see Section~\ref{sec_incl} below). For NGC\,1068 and
NGC\,3227 the value   $\Delta i>90$ is listed, meaning that the
innermost parts of the circumnuclear gaseous disks are warped so
much that the orbits are re-approaching the main plane of the
galaxy;\\
 (13) the comment, pointing to the observed composition of
the structure (H\,II---ionized, H\,I---neutral, CO---molecular
gas, s---stars) and its structure: \mbox{w---a} strong warp,
\mbox{r---a} ring, i.e. there is a hole in the center;\\
(14) references to the literature used.

 \begin{table*}
 \caption{List of galaxies with inner polar structures \label{tab1}}
\begin{turn}{90} 
\small
\begin{tabular}{l|l|l|r|r|r|r|r|r|r|r|r|r|l}\hline\hline
Name     &Type             & T    &$D$, Юяъ&$r,''$ & $r$, ъяъ&$PA_0,\degr$&$i_0,\degr$&$PA_{bar},\degr$& $PA_1,\degr$&$i_1,\degr$&$\Delta i,\degr$ & Comm. & Ref\\
  (1)    & (2)             &(3)   &(4)     &  (5)  & (6)     &  (7) &  (8) &   (9)    &  (10)  & (11)  & (12)     & (13)& (14) \\
\hline
Arp~220   &   S?            & 9.3  & 81.3   &0.3    & 0.12    & 40   & 40   &  --      & --     & --    & 90$^*$   & CO, w    &   \cite{arp220}                                   \\
IC 1548  &   S0            & -4.0 & 85.1   &1.5    & 0.62    & 78   & 59   &  --      & 349    & --    & --       & HII      &   \cite{SilAfan2008}                            \\
IC 1689  &   S0?           & -2.0 & 67.6   &10     & 3.28    & 164  & 90   &  --      & 74     & 30    & 90       & HII+s, r &   \cite{Hagen1997}                                \\
M 31     &   SA(s)b        &  3.0 & 0.78   & 180   & 0.70    & 35   & 77   &  --      & 325    & 40    & 88       & CO+HII, r&  \cite{MelchiorCombes2011}                        \\
Mrk 33   &   Im pec?       &  9.9 & 24.2   &12     & 1.41    & 116  & 59   &  --      & 163    & 47    & 39, 86   & HII+s    &  \cite{Moiseev2011}                               \\
Mrk 370  &   pec?          &  0.0 & 12.0   &11     & 0.61    & 346  & 45   &  --      & 260    & --    & $55-70^*$& HII, w   & \cite{Moiseev2011}                                \\
NGC 253  &   SAB(s)c       &  5.1 & 3.4    &5      & 0.083   & 230  & 79   &  --      & 324    &$>60$  & $78-90$  & HII, r   &\cite{AnantharamaiahGoss1996}                      \\
NGC 474  &   SA0$^0$(s)    & -2.0 & 32.7   &$>10$  &$>1.65$  & 330  & 26   &  --      & 256    & --    & --       & HII, r?  &\cite{Sauron_V},                         \\
                &                     &             &           &            &                &         &         &           &           &          &           &            & \citet{Sauron_XII} \\
NGC 1068 &   (R)SA(rs)b    &  3.0 & 16.1   &2      & 0.16    & 278  & 40   &  48      &  --    & --    & $>90^*$  &CO, w     &   \cite{Schinnerer1068}                           \\
NGC 2217 &   (R)SB0$^+$(rs)& -0.6 & 20.7   &  3    &0.30     & 6    & 21   &  111     & 20     &  --   &  $90^*$  & HII, w   &\cite{Bettoni1990}                                 \\
NGC 2655 &   SAB0/a(s)     &  0.1 & 24.2   &  $>15$&$>1.76$  & 85   &  54  &  90      &  20    &  --   &  --      & HI+HII, r   & \cite{SilAfan2004}        \\
                &                     &             &           &            &                &         &         &           &           &          &           &             &   \cite{SAURON_AGN} \\
                &                     &             &           &            &                &         &         &           &           &          &           &             &    \cite{Sparke08} \\
NGC 2681 & (R$'$)SAB0/a(rs)&  0.4 & 13.2   &  5    &0.32     & 148  & 25   &  25      & 90     & --    & --       & HII+s    & \cite{Moiseev2004}                                \\
NGC 2732 &   S0            & -2.0 & 32.4   &5      & 0.78    & 67   &  90  &  --      &  351   &$30-70$&  $77-83$ & HII      & \cite{SilAfan2004}                                \\
NGC 2768 &   E6?           & -4.5 & 23.6   & 16    & 1.83    & 95   &  90  &  --      &  347   &$30-60$&  $74-81$ & HII+CO   & \cite{Fried1994}          \\
                &                     &             &           &            &                &         &         &           &           &          &           &             &  \cite{SilAfan2004}          \\
                &                     &             &           &            &                &         &         &           &           &          &           &             &  \cite{Crocker2008}          \\ 
NGC 2787 &   SB0$^+$(r)    & -1.0 & 7.5    &6      & 0.22    & 109  &  62  &  149     &  72    &   50  &  32,76   & HII      & \cite{SilAfan2004}                                \\
NGC 2841 &   SA(r)b?       &  3.0 & 12.6   &  5    & 0.31    & 150  &  65  &154       &  68    &  --   &    --    & HII      &  \cite{Silchenko1997}                \\
                &                     &             &           &            &                &         &         &           &           &          &           &             &\cite{AfanaSil1999}                \\
NGC 2855 &   (R)SA0/a(rs)  & -0.2 & 26.5   &  4    &  0.51   & 117  & 42   &  --      &  --  &   --    &    $73^*$& HII, w   &  \cite{Coccato2007}                               \\
NGC 2911 &   SA0(s)? pec   & -2.0 & 47.0   & 4     & 0.91    & 140  & 56   &   --     &  63    &   75  &  71,88   & HII+s    & \cite{SilAfan2004}                                \\
NGC 3227 &   SAB(s)a pec   &  1.5 & 18.3   &0.9    & 0.08    & 158  & 56   &  138     &  --    & --    & $>90^*$  &CO, w     &   \cite{Schinnerer3227}                           \\
NGC 3368 &   SAB(rs)ab     &  2.2 &  13.7  &3      & 0.20    & 135  & 48   &  125     & 35     & --    & --       & HII      & \cite{Silchenko2003}                 \\
               &                     &             &           &            &                &         &         &           &           &          &           &             &  \cite{Moiseev2004}                 \\
NGC 3379 &   E1            & -4.8 &  14.3  &3      & 0.21    & 253  & 40   &  --      & 296    & --    & --       & HII+s    &\cite{Silchenko2003}           \\
                &                     &             &           &            &                &         &         &           &           &          &           &             &\cite{Sauron_V}           \\
                &                     &             &           &            &                &         &         &           &           &          &           &             &\cite{Sauron_XII}           \\
NGC 3384 &   SB0$^-$(s)?   & -2.7 &  13.7  &5      & 0.33    & 55   & 57   &  132     & 226    & --    & --       & s        & \cite{Silchenko2003}                              \\
NGC 3414 &   S0 pec        & -2.0 & 25.2   & 9     & 1.10    & 179  & 52   &   --     &  68    & $>60$ &   56, 86 & HII      & \cite{SilAfan2004}                                \\
NGC 3599 &   SA0?          & -2.0 & 20.3   & $>7$  & 0.69    & 47   & 28   & --       & $335-290$& --  & $40-60^*$& HII, w, r& \cite{SilMois2010}                                \\
NGC 3607 &   SA0$^0$(s)?   & -3.2 & 22.8   &  2    &0.22     & 300  & 34   & --       & 320    & --     &   --     & HII      &  \cite{AfanaSil2007}                              \\
NGC 3608 &   E2            & -4.8 & 22.9   & 4     &0.44     & 255  & 47   & --       & 195    & --    &   --     & HII      &  \cite{AfanaSil2007}                              \\
NGC 3626 &   (R)SA0$^+$(rs)& -0.9 & 20.0   &  4    & 0.39    & 341  &  32  & --       & 190    & --    & $58,87^*$& HII      &  \cite{SilMois2010}                               \\
NGC 4100 & (R$'$)SA(rs)bc  &  4.1 & 19.7   & 12    & 1.14    & 346  & 73   &   --     & 358    &  60   & $25,55^*$& HII      & \cite{Fridman2005}                                \\
NGC 4111 &   SA0$^+$(r)?   & -1.3 & 14.9   & $>8$  &$>0.59$  & 150  & 84   &   --     &  60    &       &   $84-90$& HII      & \cite{SilAfan2004}                                \\
NGC 4233 &   S0$^0$        & -2.0 & 35.1   & 7     & 1.19    & 176  & 87   &  200     &  76    &   90  &  --    80& HII      & \cite{SilAfan2004}                                \\
NGC 4424 &   SB(s)a?       &  1.0 & 16.8   &  3    &  0.21   & 95   & 63   &  --      &  --    & --    &   --     & HII      &  \cite{Coccato2005}                               \\
NGC 4548 &   SB(rs)b       &  3.1 & 15.6   & 3     &0.23     &145   &  37  &  110     & 236    & --    &   --     & HII      &  \cite{Silchenko2002}                             \\
\hline
\end{tabular}
\end{turn}
 \end{table*}

\setcounter{table}{0}
\begin{table*}
 \caption{(continue)}
\begin{turn}{90} 
\small
\begin{tabular}{l|l|l|r|r|r|r|r|r|r|r|r|r|l}\hline\hline
Name     &Type             & T    &$D$, Юяъ&$r,''$ & $r$, ъяъ&$PA_0,\degr$&$i_0,\degr$&$PA_{bar},\degr$& $PA_1,\degr$&$i_1,\degr$&$\Delta i,\degr$ & Comm. & Ref\\
  (1)    & (2)             &(3)   &(4)     &  (5)  & (6)     &  (7) &  (8) &   (9)    &  (10)  & (11)  & (12)     & (13)& (14) \\
\hline
NGC 4552 &   E0-1          & -4.6 & 15.6   &5      & 0.38    & 110  & 14   &  --      & 32     & --    & --       & HII      &\cite{Sauron_V},                 \\
                &                     &             &           &            &                &         &         &           &           &          &           &             &\cite{Sauron_XII}                 \\
NGC 4579 &   SAB(rs)b      &  2.9 & 23.6   & 18    & 2.06    & 96   & 39   &   58     &  154   & --    &   --     & HII      &  \cite{SAURON_AGN}                                \\
NGC 4672 &   SA(s)a pec    &  1.1 & 45.7   &  6    & 1.32    & 46   & 90   &   --     &  134   & --    &   $88-90$& s        &  \cite{Bertola2000}                  \\
               &                     &             &           &            &                &         &         &           &           &          &           &             & \cite{Pizzella2001}                  \\
NGC 4698 &   SA(s)ab       &  1.7 & 16.1   &  5    & 0.39    & 170  & 74   &   --     & --     & --    &   --     & HII+s    &  \cite{Bertola2000}                               \\
NGC 4941 &   (R)SAB(r)ab?  &  2.1 & 21.2   &  2    &  0.21   & 15   & 37   &   0      & --     & --    &   --     & HII      &  \cite{Coccato2005}                               \\
NGC 5014 &   Sa?           &  1.4 & 19.5   &22     & 2.08    & 100  & 81   &  --      & 47     & 80    & $52-56$  & HI,r     &\cite{Noordermeer2005}                 \\
                &                     &             &           &            &                &         &         &           &           &          &           &             &\cite{Moiseev_SDSS}                 \\
NGC 5198 &   E1-2?         & -4.8 & 39.3   &4      & 0.76    & 14   & 49   &  --      & 32     & --    & --       & HII+s    &\cite{Sauron_V}                         \\
                &                     &             &           &            &                &         &         &           &           &          &           &             &\cite{Sauron_XII}                         \\
NGC 5850 &   SB(r)b        &  3.1 & 38.7   &  6    &1.13     & 335  & 37   &  115     & 35     &  --   &  --      & HII      & \cite{Moiseev2004}                                     \\
NGC 6340 &   SA0/a(s)      &  0.4 & 22.4   & 12    & 1.30    & 70   &20    &85        &  330   &$40-60$& $40,65^*$& HII, w   &  \cite{Silchenko_N6340}                 \\
                &                     &             &           &            &                &         &         &           &           &          &           &             & \cite{Chil2009}                 \\
NGC 7049 &   SA0$^0$(s)    & -1.9 & 29.9   &  5    & 0.73    & 58   & 60   &  --      &  --    &  --   &    $89^*$& HII      &  \cite{Coccato2007}                               \\
NGC 7217 &   (R)SA(r)ab    &  2.5 & 16.7   & 3     &0.24     &268   &  30  &  60      & 329    & --    &    --    & HII+s    &\cite{Sil_N7217}\\
                &                     &             &           &            &                &         &         &           &           &          &           &             &\cite{SilchenkoMoiseev2006}\\
                &                     &             &           &            &                &         &         &           &           &          &           &             &\cite{Silchenko2011}\\
NGC 7280 &   SAB0$^+$(r)   & -1.3 & 28.2   & 2     &0.27     &258   &  52  &  60      & 9      &$>80$  &  68-80   & HII      &  \cite{AfanaSil_N7280}              \\
                &                     &             &           &            &                &         &         &           &           &          &           &             &  \cite{Silchenko2005}              \\
NGC 7468 &   E3? pec       &  9.0 & 31.8   &   6   & 0.92    & 180  & 45   &    --    & 120    &   60  &  49,87   & HII      &  \cite{Shalyapina2004}                            \\
NGC 7742 &   SA(r)b        &  2.8 & 24.7   &       &0.36     &128   &  9   &  --      &  130   &$>35$  & $>26$    & HII      & \cite{SilchenkoMoiseev2006}                       \\
UGC 5600 &   S0?           & -1.8 &  44.7  & 10    &2.17     &182   &  50  &  --      & 260    &  60   & $65,80^*$& HII, w, r&  \cite{Shalyapina2002,Shalyapina2007}             \\
\hline
\end{tabular}
\end{turn}
 \end{table*}

In total, Table~\ref{tab1} lists 47 galaxies, for which
there exist strong arguments that in their inner
regions\footnote{By inner regions we mean the scales smaller than
or comparable with the characteristic scales of a bulge, or an
external disk.} a substantial part of the emitting matter steadily
rotates in the plane, strongly inclined to the plane the main
disk. As a rule, such a dynamic configuration is directly stated
by the authors of papers, referred to in the last column of
Table~\ref{tab1}. In the case of other objects, we believe
that the presence of polar (or highly inclined) orbits is the most
reasonable explanation of the observed circumnuclear kinematics.
To make this conclusion we must have a velocity field, obtained
with a sufficiently high spatial resolution. Historically, the
first technique was to make several spectral sections with a long
slit, while more reliable results can be obtained with the
panoramic (3D, integral-field) spectroscopy in the optical range,
or using radio interferometry in the molecular gas lines. An
interesting example is the NGC\,253 galaxy, in which an IPS was
revealed from the observations in the radio recombination
H92$\alpha$ line~\citep{AnantharamaiahGoss1996}.

In~\citet{Sauron_XII} the terms ``kinematic twist'' and
``kinematically decoupled component'' identify the cases of
significant (exceeding \mbox{$10\degr$--$20\degr$}) variations of
\pak with increasing distance from the center in the observed
line-of-sight velocity field. Note that  such a mismatch should
not always be associated with rotation in orbits that lie outside
the galactic disk plane. The twist of the \pak can also be related
with the non-circular motions in the plane of the galaxy caused by
the  non-spherical potential of the central bar or triaxial bulge.
Fortunately, comparing the velocity field with the results of
isophote analysis of the galaxy  images, we can understand what
type of motion takes place (see the discussion and references
in~\citet{Moiseev2004}). In addition, the triaxial
potential and the inclined disk should manifest themselves in
different ways in the distribution of the line-of-sight velocities
along the major and minor axes of the
galaxy~\citep{Corsini2003}. However, the absence of the
line-of-sight velocity gradient along the minor axis is not a
sufficient criterion of the presence of an IPS. Hence, our list
lacks plenty of candidates from~\citet{Coccato2004}. On the
other hand, our sample includes most of the early-type galaxies
from the SAURON survey, for which the difference between the \pak,
determined by the velocity fields of gas and stars, is in  excess
of $30\degr$. The only exception were the objects where the disk
of ionized gas extends beyond the edge of the spectrograph's field
of view, obviously, being an internal part of the  polar
structures, observed in neutral hydrogen far beyond the stellar
disk of the galaxy. This applies to NGC\,2685---the prototype of
the classical PRG, and a galaxy with an external UV-ring,
NGC\,4262~(\citet{Bettoni2010},  also see \citet{LeoII_HI} for the HI map).

In addition to the perturbing effect produced by the triaxial
potential, a sharp change in the direction of the line-of-sight
velocity gradient can be also due to the radial gas flows,
triggered by the central burst of star formation or by a jet from
the active nucleus. Thus,  \citet{Coccato2004}, based on the long-slit
spectroscopy data suspected the existence of an IPS in NGC\,6810.
However, the subsequent studies have shown that  the central
kinematics of gas there is determined by the starburst superwind  \citep{N6810_wind}. For a similar reason we have
excluded from consideration a number of known galaxies with
ionization cones, in which the motions of ionized gas in the
central kiloparsec region are likely to be caused by the activity
of the nucleus, despite the fact that a number of authors have
found inclined disks here \mbox{(Mrk\,3, \citet{MRK3}},
\mbox{NGC\,5252, \citet{N5252}}, etc.).

Note that the existence of a twist of \pak is not strictly
necessary. For example, in NGC\,3607 and NGC\,7742, visible almost
face-on, the kinematic position angles for the external and
internal regions are almost identical \mbox{($PA_1\approx PA_0$)}.
However, the amplitude of the line-of-sight velocity  of gas,
observed in the center is so high that the most reasonable
explanation for this is rotation of the disk, strongly inclined to
the line of sight. The case of an elliptical galaxy NGC\,5198 is
challenging. This object, according to the velocity fields,
presented in~\citet{Sauron_V,Sauron_XII} has two
polar structures---an internal stellar  \mbox{($r<2''$)}  and a
more extended gaseous \mbox{($r<5''$)} structure, not coinciding
with each other. At that, the position angle of the gaseous disk
almost coincides with the $PA_{\rm kin}$ of the external regions
of the stellar velocity field, however it looks that the
counter-rotation occurs. But as the observed amplitude of the
ionized gas rotation curve is very large, it is clearly outside
the main plane of the stellar spheroid, in which the outermost gas
is rotating.

\subsection{Galaxies, not Included in the List\label{nolist}}

Compiling the list we have tried to review the largest possible
number of observational papers on this issue. Unfortunately, the
abstracts and conclusions do not always contain information on the
presence of polar structures. {\it Therefore, we apologize in advance
to the authors of papers that we may have missed.} At the same
time, our sample does not include some known candidates for the
presence of IPSs. Besides the already mentioned galaxies
possessing extended polar disks and active nuclei with ionization
cones, these are objects on which, in our opinion, the available
data is insufficient or too controversial. These include:
NGC\,3672, mentioned in the introduction (the galaxy where the
existence of an IPS was first reported), as well as most of the
other galaxies with the remarkable line-of-sight velocity gradient
along the minor axis from the list of~\cite{Coccato2004};
the kinematically decoupled nucleus in NGC\,4150, known from the
SAURON/OASIS data~\cite{Sauron_VII}; NGC\,7332, although
having some indications of the presence of an inclined
disk~\cite{Silchenko2005}, its velocity field has a very
irregular shape; NGC\,524, where a polar disk was
suspected~\citep{Silchenko_N6340}, but the new data reveal
a more complex structure consisting of two counter-rotating
disks~\cite{Katkov2011}; NGC\,3367, in which according
to~\cite{Gabbasov2009} there is a reversal of internal
isovels in the line-of-sight velocity field, but the
interpretation is difficult; M83, which was suspected in the
presence of an IPS  from the morphology of dust
lanes~\cite{SofueWakamatsu1994}, but the follow-up studies
of perinuclear kinematics did not confirm this hypothesis.

A recent paper~\cite{ATLAS3D_X} provides the data on the
kinematic parameters of early-type galaxies within the
ATLAS$^{\mbox{3D}}$ surveys. Their lists reveal a dozen more
galaxies in which the difference of position angles, measured by
the velocity fields of gas and stars exceeds
\mbox{$40\degr$--$50\degr$}.  However, we can not be sure that we
are dealing with the inner polar disks, since the authors do not
give neither any \pak$(r)$ dependencies, nor the velocity fields
of ionized gas. We have also eliminated from our consideration the
objects in which the \pak coincides with the photometric major
axis in the inner regions, and with the minor axis in the external
regions (NGC\,4365, NGC\,4406, etc.), what is most likely related
with the triaxiality of these elliptical galaxies (see the
survey~\cite{deZeeuwFranx1991})

\section{STATISTICAL PROPERTIES\label{sec_stat}}

\subsection{General Remarks}

The compiled list confirms the idea that the inner polar
structures are a very common
phenomenon~\cite{Coccato2004}. Indeed, the number of known
galaxies, containing IPSs is one and a half times greater than the
total number of galaxies with kinematically confirmed external
polar rings (about 30 objects, see~\cite{Moiseev_SDSS}).
Unlike the case of ``classical'' PRGs, among the IPSs only
relatively close objects are as yet available for detailed
observations: most of galaxies from Table~\ref{tab1} are
located closer than \mbox{30--40~Mpc} from us, including three,
belonging to the Local Volume  ($D<10$~Mpc). When we mean the
PRGs, it is reasonable to talk about the rings, even if they are
rather wide, on the other hand,  considering the IPSs, we are as a
rule (in 39 out of 47, i.e. 83\% of cases)  talking about the disk
geometry in which the inner diameter is negligible in comparison
with the outer one. Unfortunately, we do not possess enough
observational data of high spatial resolution to refine their
detailed inner structure. As follows from
Table~\ref{tab1}, only the gaseous disks and rings are
mainly found, the list contains only one galaxy (NGC\,3384) with a
purely stellar polar disk, and six more structures are marked as
stellar-gaseous. The selection effect is present here, since
methodologically it is much easier to identify the emission lines
of ionized or molecular gas than to separate the absorption
spectrum, observed along the line of sight, into stars, belonging
to the bulge, the polar disk and the main disk of the galaxy. It
is obvious that quite a few described IPSs are indeed
stellar-gaseous, just that it is problematic to detect the
kinematic manifestation of stars. For example,  the stellar polar
disk in NGC\,7217, identified by its kinematic properties is
remarkably more compact than the gaseous
one~\cite{SilchenkoMoiseev2006}. The ionization of gas in
the IPSs can (at least partially) be explained by the ongoing star
formation.

\subsection{Radial scales\label{sec_size}}   

\begin{figure*}[tbp!!!]
\centerline{\includegraphics[width=8.cm]{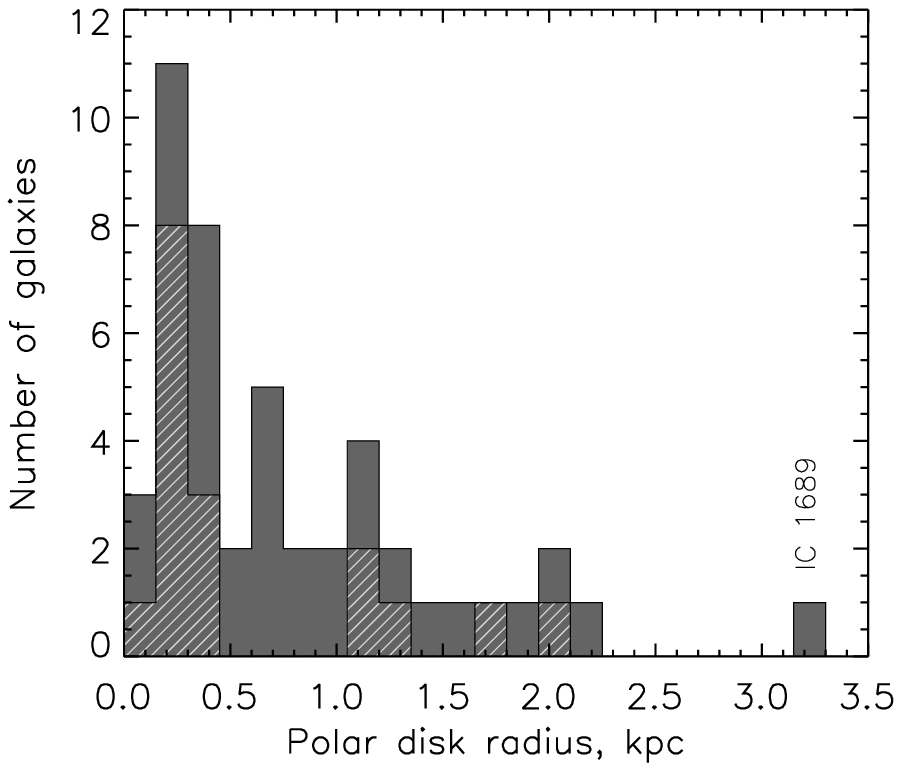}\hspace{10mm}
\includegraphics[width=8.cm]{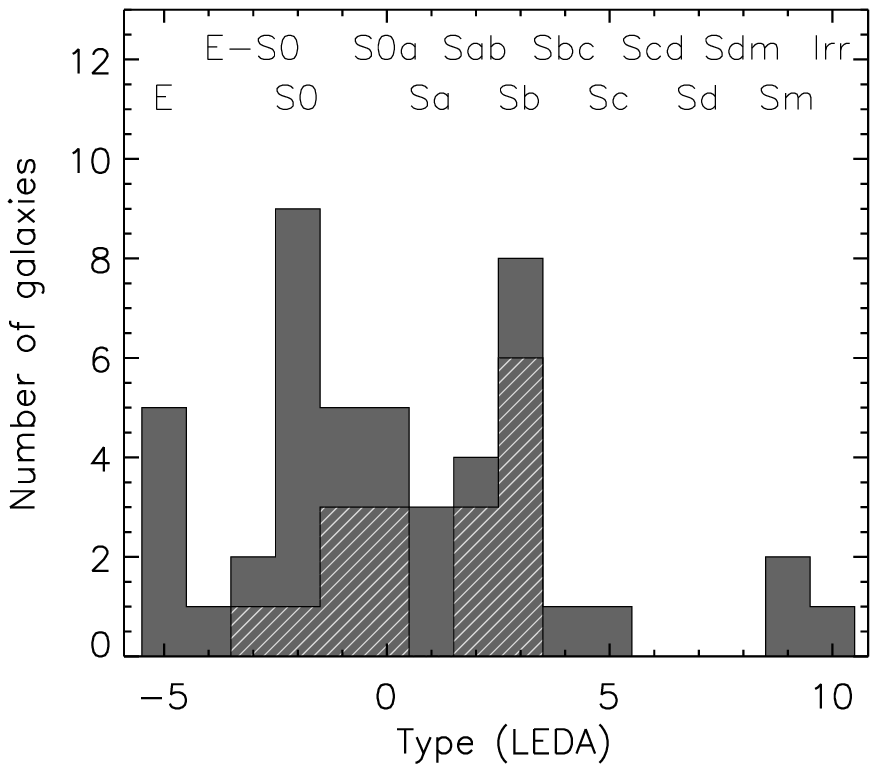}}
\caption{The histograms of distribution of the internal polar
structures by sizes (left) and morphological types of galaxies
(right). All the sample are marked in gray, while the galaxies
with bars and triaxial bulges are shaded.\label{fig_1}}
\end{figure*}

Figure~\ref{fig_1} shows the histogram of distribution of
the sizes of internal polar structures. It it clear that with
respect to them it would be fair to use the term the ``central
kiloparsec'': the average median radius is about $600$~pc, 85\% of
all IPSs are sized below 1.5~kpc. This compactness is likely due
to the fact that for a stable existence of polar orbits a
stabilizing factor is required, i.e. a spheroidal or triaxial
potential. For the classical PRGs it is the gravitational
potential of dark halos, and for the internal structures---the
potential of a bulge or a bar (see Section~\ref{sec_bar}
below). This may practically explain the lack of known polar
structures of intermediate size, with \mbox{$r=2$--$10$~kpc}.
Evidently, at these scales, the differential precession that
occurs under the influence of gravitational potential of the
stellar disk leads to a catastrophically fast decrease of orbit
inclinations and their ``fallout onto the disk.'' Note that even
in the IPS in IC\,1689---the largest in our list---the radius is
smaller than the effective radius of the bulge (\mbox{$r_e=4$~kpc}
according to~\cite{Reshetnikov1995}). It is possible that
it is exactly the picture, observed in NGC\,7743---a lenticular
galaxy where the entire ionized gas at \mbox{$r=1.5$--$5.4$~kpc}
is located in the plane, inclined by $34\degr$ or $77\degr$ with
respect to the stellar disk~\cite{Katkov2011_N7743}.

The dip in the distribution for $r<200$~pc is apparently caused by
the limited spatial resolution of most of observations, since this
scale at $D=30$~Mpc corresponds to an angular size of about
$2\arcsec$.

\subsection{Morphological Types\label{sec_type}} 

We know that the outer polar rings are as a \linebreak rule
observed around the early-type galaxies, \linebreak
E/S0~\mbox{\cite{Whitmore1991,Reshetnikov2011}}.
One of the explanations for this is that these galaxies have poor
own internal gas reserves, hence the gas clouds at the polar
orbits do not undergo any collisions with the gas in the main
galactic plane. To what extent is this true for the internal polar
structures? The estimates made by \cite{Coccato2004} have shown that the remarkable
line-of-sight velocity gradient of gas along the minor axes of
galaxies is predominantly observed in the S0 galaxies and
early-type spirals, which may indicate that there is a relation
between this phenomenon and the presence of a vigorous bulge.
However, the authors themselves noted that the velocity gradients
along the minor axis are not always due to rotation in the polar
plane.

If we consider only the galaxies with confirmed IPSs, the
distribution by morphological type becomes broader
(Fig.~\ref{fig_1}, right). The median average here is
$T=0$, i.e. only a half of all galaxies belongs to the S0a type
and earlier. And nearly a third of them are the \mbox{Sb-type}
objects and later, even including a few Sm and Irr galaxies,
commonly possessing small  bulges. Note that the actual number of
late-type galaxies must be underestimated by the selection effect,
since the large surveys of kinematics of the circumnuclear regions
with the MPFS and SAURON integral-field-spectrographs were focused
primarily on the E/S0 galaxies.

\begin{figure*}[tbp!!!]
\centerline{\includegraphics[width=8.cm]{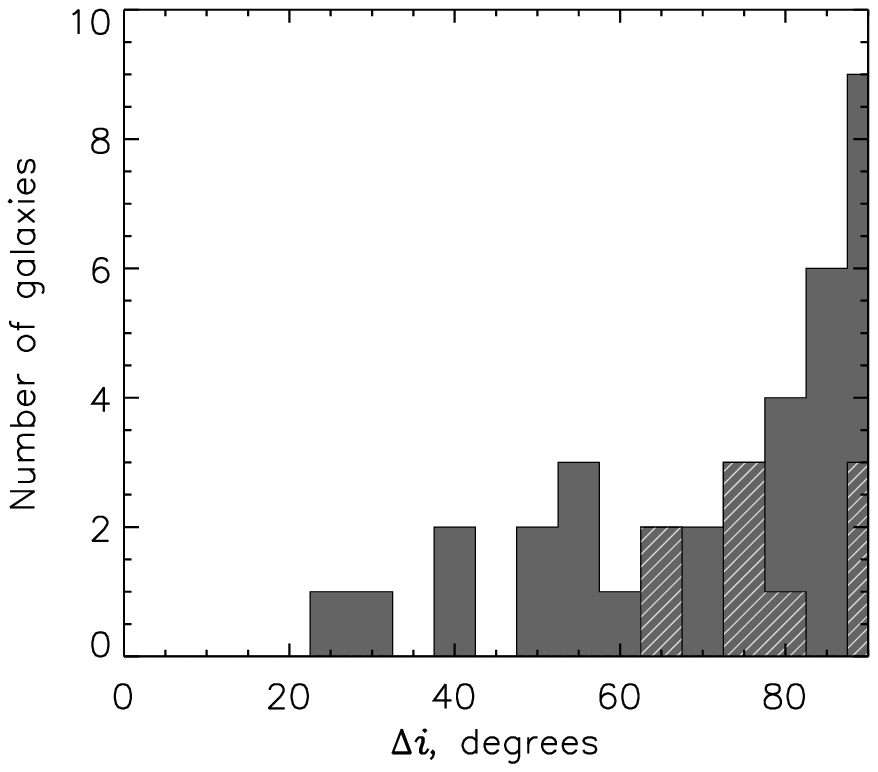}\hspace{10mm}
\includegraphics[width=8.cm]{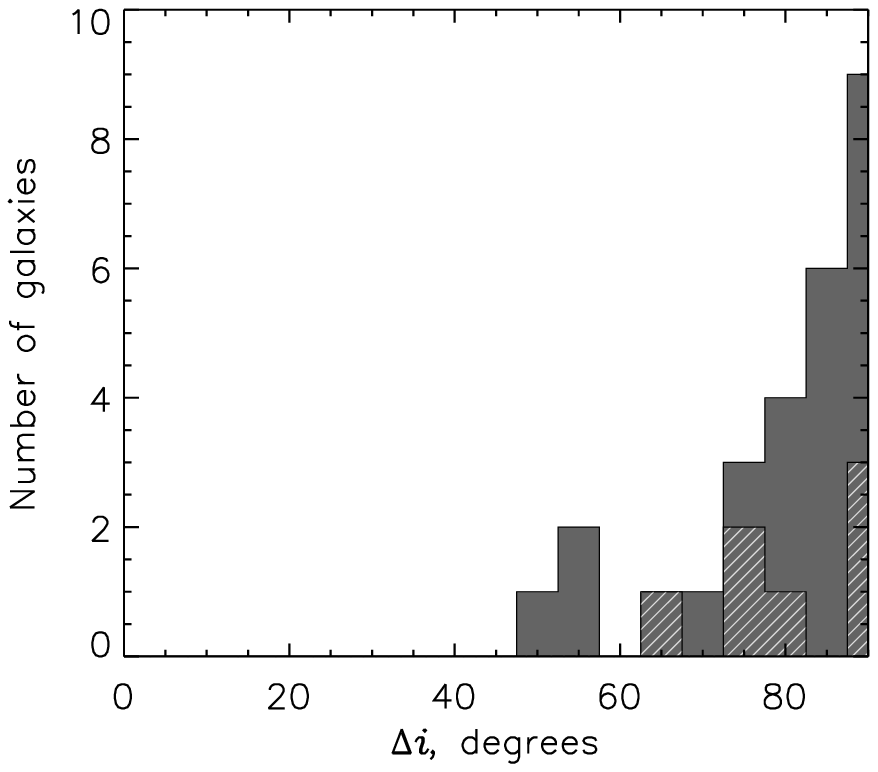}}
\caption{The distribution by the inclination angle of inner rings
and disks to the main galactic disk. The left plot: the entire
possible range of solutions for   $\Delta i$ from
(\ref{eq1}), the right plot: in the case of two solutions
the greater was taken.  All the galaxies are marked in gray, while
the galaxies with bars and triaxial bulges are shaded.
\label{fig_2}}
\end{figure*}

Therefore, we can make a preliminary conclusion that for the
existence of a circumnuclear polar or inclined disk, the
morphological type of a  galaxy is less significant than for the
PRG. Apparently, the effect of collision of  gas at the polar
orbits with the gas of the main disk is not critical for the
formation of these structures. It is possible that the forming
inner polar disk has time to preliminarily ``sweep up'' the region
of the central kiloparsec. In any case, the examples of strongly
warped IPSs are known, when the gas near the nucleus rotates in
the polar plane, and with increasing distance from the center, the
orbits fall into the plane of the galaxy (see
Section~\ref{sec_incl} below). It is possible that in some
cases the inclined orbits are occupied by the gas from the main
plane of the galaxy under the effect of the gravitational
potential of the bar (Section~\ref{sec_bar}). Note,
however, that the histograms in Fig.~\ref{fig_1} do not
show any significant differences in the distribution of barred
galaxies.

\subsection{Inclined or Polar?\label{sec_incl}}  

Using the term ``polar'' with respect to the IPSs, one has to
remember that is not always possible to accurately measure the
inclination of the plane of the inner disk to the outer. It is
easy to show~(\cite{Moiseev2008}) that this angle $\Delta
i$ is expressed by the relation:

\begin{equation}
\begin{array}{r}
\cos \Delta i =\pm\cos (PA_0-PA_1) \sin i_0 \sin i_1  \\ +\cos i_0
\cos i_1. \label{eq1}
\end{array}
\end{equation}

Most often we only know  the parameters of orientation of the
outer disk ($PA_0$, $i_0$) and the direction of the major axis of
the inner structure $PA_1$. However, to determine the inclination
angle $i_1$ of the orbits to the line of sight from the observed
kinematics within the model of circular rotation, a detailed
velocity field with a large number of independent points is
required. Moreover, a stable solution can usually be obtained only
for a notable inclination of the disk to the line of sight
\mbox{($i_1>30$--$40\degr$)}. An only exception is a case of a
purely polar disk in a galaxy, seen edge-on
(\mbox{$PA_0=PA_1+90\degr$}, $i_0=90\degr$). The uncertainty with
the sign of the first term in~(\ref{eq1}) is due to the
fact that $PA$ and $i$  do not fully characterize the position of
the plane with respect to the observer---we also need to know the
direction of the moment of rotation, i.e. which side of the disk
is nearer to us, and which is farther. Hence, for a number of
galaxies with $i_0<90\degr$ Table~\ref{tab1} gives both
possible alternate solutions of~(\ref{eq1}). One of the
few exceptions is the polar ring in the Andromeda galaxy (M\,31).
Its relatively large angular scale has allowed the authors
of~\cite{MelchiorCombes2011} to precisely understand how
it is oriented with respect to the galactic disk.

The angle $\Delta i$ was estimated for 27 objects, which is
slightly over a half of the entire sample. The histograms
presented in Fig.~\ref{fig_2} resemble much the
distribution by the same parameter of the outer polar rings
from~\cite{Whitmore1991}. Notwithstanding the above
uncertainty with the estimation of $\Delta i$, most of the outer
inclined structures turn out to be truly polar, i.e. perpendicular
to the outer disks of galaxies. Thus, $\Delta i
>70\degr$ in 23 out of 27 (i.e.  85\%) cases.

On the other hand, even if we choose between the two solutions for
$\Delta i$ the option, closest to $90\degr$
(Fig.~\ref{fig_2}, right), there are still IPSs located at
a more moderate angle $\Delta i$ equal to $50\degr$--$60\degr$:
NGC\,3599,  NGC\,5014 and NGC\,6340. Such structures are unlikely
to be stable: they have to  relatively quickly fall into the
galactic plane under the effect of differential precession. An
indirect indication of this are the observed warps of gaseous
disks in NGC\,3599, and NGC\,6340. Note that similar warps are
found in seven galaxies of the sample. In many cases, the authors
can construct a detailed spatial model of such warped disks,
reproducing not only the kinematics, but also the brightness
distribution of ionized (NGC\,2855, and NGC\,7049) or molecular
(Arp\,220, NGC\,1068, NGC\,3227) gas (see references in
Table~\ref{tab1}).

\subsection{Bars and Triaxial Bulges\label{sec_bar}} 

The relationships of internal polar structures with the
nonaxisymmetric gravitational potential have been widely
discussed in the literature, starting from the
paper~\cite{Bettoni1990}, for the first time describing
the  case of a circumnuclear polar disk in a barred galaxy
NGC\,2217, oriented virtually perpendicular to the major axis of
the bar. Further on,  similar arrangements of IPSs were found in
many other barred galaxies, listed in Table~\ref{tab1}.
It was repeatedly noted that such an arrangement of the polar disk
along the smallest section of the bar---i.e. in one of the main
planes of a triaxial potential---has to be stable. At that, it is
sufficient to have a non-spherical (triaxial) bulge in a galaxy
instead of a large-scale
bar~\citep{Silchenko_N6340,AfanaSil1999,Corsini2003}.

\begin{figure}[tbp!!!]
\centerline{\includegraphics[width=8.0cm]{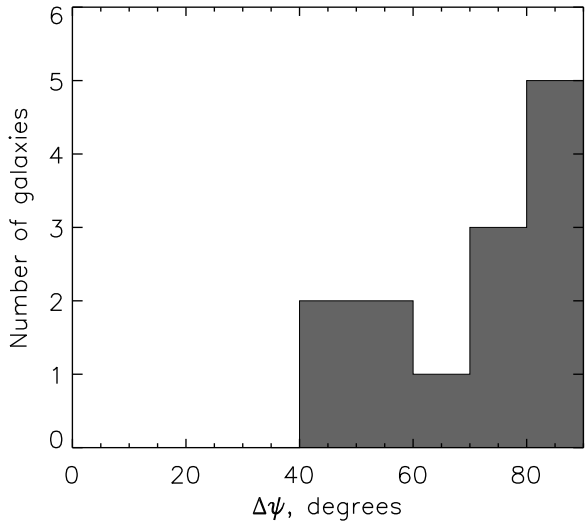}}
\caption{The distributions by the angle between the major axis of
the polar disk and the bar (or the triaxial bulge).
\label{fig_3}}
\end{figure}

Figure~\ref{fig_3} presents  the distribution of all our
sample galaxies by the angle $\Delta \psi$  between the major
axis of the nonaxisymmetric stellar structure (a bar or a
triaxial bulge) and the major axis of the IPS (projected on the
galactic plane):

\begin{equation}
\begin{array}{r}
\Delta \psi = \arctan \displaystyle\frac{\sin(PA_{\rm
bar}-PA_0)\cos i_0}{\cos (PA_{\rm bar}-PA_0)} \\ \\
-\arctan \displaystyle\frac{\sin (PA_1-PA_0)\cos i_0}{\cos
(PA_1-PA_0)}.
\end{array}
\end{equation}

Despite the relatively small statistics available, the tendency of
polar structures to align orthogonally to the major axis of the
bar ($\Delta \psi=90\degr$) is clearly visible. So far, the
literature has detailed calculations of the formation of such
inner disks, for instance, via the capture of the outer gas clouds
with the corresponding direction of the orbital moment. Usually
the authors use a star-dynamic analogy with polar or warped disks,
observed in the global triaxial potential of elliptical galaxies
(see discussion
 in~\citet{Corsini2003} and \citet{SilAfan2004}). In any
case, it is most likely that triaxiality of the potential is
responsible for the stabilization of the internal polar disks in
the circumnuclear regions of   elliptical galaxies (NGC\,3608,
NGC\,4552, etc.). Besides the scenario with the capture of gas
clouds with the corresponding orbital moment, the
studies~\cite{SofueWakamatsu1994}
and~\cite{Friedli1993} are often quoted. The authors of
the former paper made an assumption that the internal polar disk
formed under action of the gravitational potential of the bar on
the warped gaseous disk of the galaxy. This leads to the loss of
angular momentum in the azimuthal plane and its conservation in
the polar plane. \citet{Friedli1993} have
demonstrated with the aid of a three-dimensional numerical model
that if a part of gas in the disk of a galaxy initially rotated in
the opposite direction with respect to the rest of the disk, in
the process of secular evolution of the bar the gas clouds occupy
stable orbits, strongly inclined to the galactic plane.
Olga Sil'chenko and her colleagues have repeatedly emphasized in
their numerous papers the cases of simultaneous observations of
polar disks in the inner regions of galaxies and the
counter-rotation of gas--stars or stars--stars in the outer
regions. This feature is detected in 9 galaxies of our list (see
the following section). The initial counter-rotating component is
most likely the result of absorption of a dwarf companion. Note
two points, however. Firstly, in the modeling made by  \citet{Friedli1993}, the obtained disk is not polar, but inclined at about
$45\degr$ to the plane of the galaxy. Secondly, the proposed
mechanisms can obviously not be the main method of polar disk
formation, since in this case one should expect an increased
number of bars in the galaxies with IPSs. However, out of 40 disk
galaxies (S0-type and later) from our sample only 17 possess
confirmed bars or triaxial
bulges\footnote{Table~\ref{tab1} lists angle $PA_{\rm
bar}$ for all the galaxies, where we managed to find references on
the presence of an internal triaxial structure. We mainly focused
on the studies devoted to a detailed analysis of the morphology of
galaxies (including those using the images in the near-IR, optimal
for the search of bars). We believe that such an approach is more
correct than the use of a morphological classification from
RC3/NED.}, which is \mbox{$43\pm8\%$} or \mbox{$33\pm7\%$}, if we
ignore  the non-spherical bulges\footnote{Hereinafter the variance
of the binomial distribution is specified as an error.}. This is
in good agreement with the known frequency of bars in nearby
galaxies, which is about $45\%$~\citep{Aguerri2009}.

Therefore, we must conclude that although the bars have an effect
on the orientation of polar disks, the existence of a triaxial
potential is not absolutely necessary for the existence of IPSs.
Surely, the galaxies may have a certain internal triaxiality,
barely noticeable to the detached observer (see, e.g., the
discussion of this issue in~\cite{AfanaSil2007}). However,
the circumnuclear regions usually look more symmetrical than the
outer ones not only in the unbarred galaxies, but also in the
barred spirals (inside the Lindblad resonances of the bar).
In~\cite{SilMois2010}, we show that NGC\,3599 and
NGC\,3626 have an internal ``oval distortion of the disk'', which
may point to the formerly existing bars, destroyed in the process
of secular evolution or under the effect of external influence.
However, such a scenario is   unlikely to fit most of the
remaining unbarred galaxies.

\subsection{External Environment\label{sec_external}}

Another frequently debated issue is the relationship of IPSs with
the external environment of galaxies, and the processes of their
interaction. By analogy with external polar rings, it is
reasonable to expect that the circumnuclear polar structures can
as well  be formed as a result of the capture of matter (gas
clouds or a dwarf companion) with the orbital momentum, orthogonal
to the moment of rotation of the galactic disk. Typically, such
discussions considered individual cases, where there are either
clear signs of a recent interaction, or vice versa---a galaxy is
isolated and looks unperturbed by any external effects. The
examples of an explicit relation of IPSs with the environment are:
NGC\,2655~\citep{Sparke08}, where the polar disk of ionized
gas is an internal part of a strongly warped extended disk of
neutral hydrogen with pronounced tidal structures in the outer
regions, low-contrast ripples on the optical images of
NGC\,474~\citep{arp} and
NGC\,6340~\citep{Zasov2008,Chil2009},   the
existence of two kinematic H\,I components with different laws of
rotation in NGC\,3414~\citep{Morganti2006}.

\begin{table*}
\caption{Signs of effects introduced by the environment} \label{tab2}
\begin{tabular}{l|l|l|l}\hline\hline
Name      &  Optical             &  HI                    & counter-rotation            \\
(1)       &      (2)             &     (3)                &  (4)                         \\
\hline
Arp~220    & \cite{arp}           &                        &                               \\
IC 1548   &                      &                        & \cite{SilAfan2008}         \\
IC 1689   &  \cite{Reshetnikov1995}&                      &                             \\
M 31      &  \cite{M31_stream}   & \cite{M31_HI}          &                                 \\
 Mrk 33   &                      & \cite{Bravo-Alfaro2004}&                              \\
NGC 253   &  \cite{N253_tidal}   &                        &\cite{AnantharamaiahGoss1996} \\
NGC 474   & \cite{arp}           &                        &                               \\
NGC 2655  &  \cite{Sparke08}     &  \cite{Sparke08}       &                              \\
NGC 2681  & \cite{Cappellari2001}  &                        &                                \\
NGC 2768  &                      &  \cite{Morganti2006}	  &                              \\
NGC 2787  &                      &\cite{Shostak1987}      &                              \\
NGC 2841  &                      &                        &  \cite{AfanaSil1999}         \\
NGC 2855  &\cite{Corsini2002}    &                        &                               \\
NGC 3227  & \cite{arp}           &                        &                               \\
NGC 3368  &                      & \cite{LeoI_HI}         &                                 \\
NGC 3379  &                      & \cite{LeoI_HI}         &                                 \\
NGC 3384  &                      & \cite{LeoI_HI}         &                                 \\
NGC 3414  &                      &  \cite{Morganti2006}	  &  \cite{SilAfan2004}         \\
NGC 3607  &                       &  \cite{LeoII_HI}       &                             \\
NGC 3608  &                       &  \cite{LeoII_HI}       &  \cite{N3608_counter}       \\
NGC 3626  &                       &                        &  \cite{Ciri1995}            \\
NGC 4111  &                       &   \cite{ATLAS_HI}      &                      \\
NGC 4424  &   \cite{N4424_tails} &     \cite{N4424_HI}     &                             \\
NGC 4672  &  \cite{N4672_interact} &                        &                            \\
NGC 4698 &                      &   \cite{N4698_HI}      &                               \\
NGC 5014  &                       & \cite{Noordermeer2005} &                             \\
NGC 5850  &  \cite{N5850_HI}    &                        &                                 \\
NGC 6340  &\cite{Zasov2008,Chil2009}&                     &                              \\
NGC 7217  &                       &                        &  \cite{N7217_counter}             \\
NGC 7280  &                       &                        &  \cite{AfanaSil_N7280}            \\
NGC 7468  &\cite{Evstigneeva2000}&                        &                              \\
NGC 7742  &                      &                        & \cite{SilchenkoMoiseev2006}  \\
UGC 5600  & \cite{Shalyapina2007} &                        &                             \\
\hline
\end{tabular}
\end{table*}

An example of an integrated approach to the problem is a recent
paper, devoted to the statistics of the differences in the
kinematics of gas and stars in the
ATLAS$^{\mbox{3D}}$~\cite{ATLAS3D_X} survey, performed
with the SAURON  integral-field spectrograph. However, firstly, these
results only relate  to the early-type galaxies, and secondly, as
we have already noted above, not every case of  ``kinematic
misalignment''  indicates  rotation at polar or inclined orbits.

Table~\ref{tab2} contains the data on the galaxies of our
sample, for which there is evidence of a recent (on the scale of
less than 1--2~Gyr) interaction. Column (1) gives the name of the
galaxy, (2) bears references to papers, indicating interaction
with a companion or a merger of a satellite based on the optical
photometry data, (3) gives indications of tidal structures or
nearby clouds, observed in H\,I, (4) contains an indication of the
presence of a counter-rotating component in the galactic disk. In
total such evidence were collected for 33 galaxies, making up
\mbox{$70\pm7\%$} of the entire IPS sample. Such a high proportion
of galaxies with signs of recent interaction allows us to conclude
that the effects of external environment do indeed play a major
role in the formation of internal polar rings and disks.

\section{SUMMARY AND CONCLUSIONS\label{sec_conclusion}}

We compiled the list of galaxies with polar (or strongly inclined
to the main the galactic plane) disks and rings detected in their
inner regions. It is interesting to note that more than a half
($60\%$) of described structures have been discovered or
confirmed as a result of observations at the  SAO RAS \mbox{6-m}
BTA telescope using the MPFS integral-field spectrograph, or the
SCORPIO instrument in the scanning Fabry-Perot interferometer
mode. The study of statistical properties of various parameters,
characterizing these inner structures allows us to draw the
following conclusions.

\begin{enumerate}

\item The stellar-gaseous polar disks and rings are found in
the central regions of galaxies of all morphological types.

\item The vast majority of inner polar structures have the
radius of less than 1.5~kpc. This limitation can be associated
with the stabilizing role of the bulge.

\item   The innermost regions of these structures are as a rule
located in the polar plane, while farther from the nucleus we
frequently observe a warp---the orbits approaching the galactic
plane.

\item The inner polar structures are equally common
in galaxies with and without bars. At the same time, if galaxy
has a bar (or a triaxial bulge), this leads to the stabilization
of the polar disk so that its axis of rotation coincides with the
major axis of the bar.

\item Seventy percent of galaxies with inner polar structures
reveal signs of a recent interaction, pointing to the leading
role of the external environment in the formation of these
peculiar structures.

\end{enumerate}

It is as yet difficult to estimate the frequency of occurrence of
circumnuclear polar disks, since our sample is composed from very
heterogeneous sources. We can only conclude that since the number
of such galaxies is one and a half times greater than the number
of kinematically confirmed PRGs, and they are on the average
located much closer, their fraction among the fairly bright
galaxies should significantly exceed the known estimates of the
occurrence of \linebreak
PRGs~\cite{Whitmore1990,Reshetnikov2011} and
amount to at least \mbox{3--5\%}. It is possible that a careful
analysis of kinematics of all the galaxies from the
ATLAS$^{\mbox{3D}}$ survey, instead of the early-type galaxies
alone (as in~\cite{ATLAS3D_X}) will allow to give much
more accurate estimates of the fraction of galaxies with IPSs.

In contrast to the external large-scale polar rings, in which the
numerical modelling succeeds to reproduce the main stages of
formation (see examples and references
in~\cite{BournaudCombes2003,Combes06}), such
modelling has not yet been performed for the internal structures.
However, the provided statistics  of the IPS properties indicates
that just like the classical PRGs, the vast majority of internal
polar rings and disks were formed as a result of capture of matter
from the external environment of galaxies. Moreover, in a recent
work~\cite{Silchenko2011} the authors, using  models grids
from the GalMer database, have demonstrated that as a result of
interaction of a giant S0 galaxy with a dwarf companion rich in
gas, a ring of star formation has formed in the bulge region,
strongly inclined to the plane of the galaxy. The companion mass
ratio is \mbox{$1:10$}, i.e. it corresponds to the minor merging
events. Specific conditions of interaction have to be satisfied,
the companion must initially be at the orbit with the  
 retrograde motions, while the planes of disks of both companions
have to be nearly orthogonal. We hope that the development of such
models will allow to better conceive the processes of formation of
IPSs in particular galaxies and reproduce their observed
parameters.

We should not be confused by the lack of visible signs of recent
interaction in approximately $1/3$ of galaxies from the list.
Firstly, sufficiently deep optical images and H\,I  distributions
are not available for all the galaxies. Secondly, it is possible
that some time after the interaction with a dwarf companion, the
presence of matter at  polar orbits of the inner part of the
galaxy may turn out to be the only evidence of this event. It is
therefore important to study the stability of internal polar
orbits in the real gravitational field of galaxies, comprising
the contribution of the disk, bulge, bar and halo. It would be
interesting to know whether our preliminary assumption of the
absence of stable structures of intermediate size between the
central kiloparsec region and the outer boundary of the stellar
disk is indeed true (Section \ref{sec_size}).

In principle, in some cases it is possible to form an IPS even
without the interaction with the environment. For example, it is
demonstrated
in~\cite{Schinnerer3227,Schinnerer1068} that
compact (about 100~pc) disks of molecular gas in the circumnuclear
regions of NGC\,1068 and NGC\,3227 could become strongly warped
under the effect of the ionization cone and radiation pressure
from the active nucleus. However, this scenario is clearly not
suitable for most of galaxies with more extended polar disks and
rings.

\begin{acknowledgements}
The work was carried out with the financial support of the
`Dynasty ' Foundation and the Ministry of Education and
Science of Russian Federation (state contracts no.~14.740.11.0800,
16.552.11.7028, 16.518.11.7073). Working on this paper, we used
the NASA/IPAC extragalactic data\-base (NED), managed by the Jet
Propulsion Laboratory of the California Institute of Technology
under the contract with the National Aeronautics and Space
Administration (USA). The author thanks the referee,
O.~K.~Sil'chenko for helpful comments, and A.~A.  and
P.~A.~Smirnovs for their help in editing the text of the paper.
\end{acknowledgements}


\end{document}